\documentclass{aa}
\usepackage {color}
\usepackage {epsfig}
\usepackage {amstex}
\usepackage {amssymb}
\usepackage {epic}
\usepackage {array}
\usepackage {natbib}
\bibpunct{(}{)}{;}{a}{}{,}
\definecolor{hgrau}{gray}{.75}

\def\quadfeld(#1,#2)(#3,#4)#5%
{\multiputlist(#1,#2)(#3,#4)[#5]{%
{\vector(#4,#3){3.75}},%
{\vector(#4,#3){7}},%
{\vector(#4,#3){9.75}},%
{\vector(#4,#3){12}},%
{\vector(#4,#3){13.75}},%
{\vector(#4,#3){15}},%
{\vector(#4,#3){15.75}},%
{\vector(#4,#3){16}},%
{\vector(#4,#3){15.75}},%
{\vector(#4,#3){15}},%
{\vector(#4,#3){13.75}},%
{\vector(#4,#3){12}},%
{\vector(#4,#3){9.75}},%
{\vector(#4,#3){7}},%
{\vector(#4,#3){3.75}}%
}}
\def\dreibein(#1,#2)#3#4#5#6#7%
{\put(#1,#2)%
{\put(0,0){\vector(1,0){#3}}%
\put(0,0){\vector(0,1){#3}}%
\put(0,-#3){\line(0,1){#3}}%
\put(0,0){\vector(1,1){#4}}%
\put(#3,-#5){{\small$x$}}%
\put(#4,#4){{\small$\,y$}}%
\put(-#5,#3){{\small$z$}}%
\put(-#5,-#5){{\tiny$0$}}
}}
\newcommand{\Eqref}[1]{Eq.~\eqref{#1}}
\newcommand{\Bvec}{\vec{B}}
\newcommand{\bvec}{\vec{b}}
\newcommand{\eBvec}{\vec{e}_B}
\newcommand{\Evec}{\vec{E}}
\newcommand{\jvec}{\vec{j}}
\newcommand{\uvec}{\vec{u}}
\newcommand{\Omvec}{{\boldsymbol{\Omega}}}
\newcommand{\omvec}{{\vec{\omega}}}
\newcommand{\tH}{{\text{H}}}

\newcommand{\tP}{{\text{P}}}
\newcommand{\tT}{{\text{T}}}
\newcommand{\ti}{{\text{i}}}
\newcommand{\curl}{\operatorname{curl}}
\newcommand{\dv}{\operatorname{div}}
\newcommand{\sign}{\operatorname{sign}}
\newcommand{\lapl}{\operatorname{\Delta}}
\newcommand{\xvec}{\vec{x}}
\newcommand{\exvec}{\vec{e}_x}
\newcommand{\ezvec}{\vec{e}_z}
\newcommand{\zervec}{\vec{0}}
\newcommand{\tmax}{{\text{max}}}
\newcommand{\kvec}{\vec{k}}
\newcommand{\undertilde}[1]{\mbox{$\underset{\displaystyle\hspace*{-.06em}\widetilde{}}{#1}$}}

\begin{document}
\title{Non--linear magnetic field decay in 
neutron stars --\\ 
Theory and observations}
\author{U. Geppert \and M. Rheinhardt}  
\institute{Astrophysikalisches Institut Potsdam,
 An der Sternwarte  16, 14482 Potsdam, Germany}
\titlerunning{Non--linear Magnetic Field Decay}
\date{Received \today / Accepted \today}

\abstract{
There exists both theoretical and observational evidence that the magnetic field
decay in neutron stars may proceed in a pronounced non--linear way during a certain episode
of the neutron star's life.
In the presence of a strong magnetic field the Hall--drift dominates
the field evolution in the crust and/or the superfluid core of neutron stars.
Analysing observations of $P$ and $\dot{P}$ for sufficiently old isolated pulsars
we gain strong hints for a significantly non--linear magnetic field decay.  
Under certain conditions with respect to the geometry and strength
of a large--scale magnetic background field 
an instability is shown to occur which rapidly raises small--scale magnetic field modes. 
Their growth rates increase with the background field strength and may 
reach $\sim 10^4$ times the ohmic decay rate. Consequences for the 
rotational and thermal evolution as well as for the cracking of the crust
of neutron stars are discussed.
}

\maketitle

\section{Introduction}

There is still an ongoing scientific debate whether the magnetic fields of 
pulsars decay at all \citep[see, e.g.,][ and references therein]{RF01}. Thus, the question whether this questionable decay proceeds
in a non--linear way seems to be somewhat academic. However, the neutron star
magnetic fields are the strongest known in the universe and already this very fact suggests
the idea that, if there is a field evolution at all, it should be a non--linear one.
As we will show, there exist both theoretical and observational evidences
that the field decay proceeds essentially in a non--linear way,
at least in certain regions of
the star and surely only during certain episodes of its life.

In order to be able to deal with well defined physical conditions we confine ourselves
here to
considering the evolution of {\em sufficiently old isolated neutron stars} (SOINSs).
We characterize them by the following properties:
\begin{itemize} 
\item the supernova fall--back accretion phase is 
finished and a stable density stratification $\rho(r)$ has been established;
\item the rediffusion of the magnetic field has been
completed;
\item  the rotation is slow enough so that the irradiation of gravitational waves
does not contribute to the spin down;
\item the appearance of glitches is less probable;
\item the crust is almost completely crystallized;
\item the temperature profile has become flat enough, to avert convection in the outermost
thin liquid layer and in the core region as well as the occurrence of a thermoelectric instability in the crust;
\item the electron relaxation time is already so large that the magnetization parameter
may reach values large enough to make the electron transport processes non--linear;
\item the absence of accretion, as a consequence of isolation, ensures
that there are neither external sources of angular momentum nor  
of heating;
\item therefore, there is a stable compactness ratio $M/R$ and no screening of 
the magnetic field can happen.
\end{itemize} 
All these constraints are met by the majority of neutron stars the age of which is $\gtrsim 10^5$ years.
Therefore, if any thermal and/or rotational evolution of SOINSs is observed beyond that 
predicted when assuming a constant magnetic field, it is with a high probability 
caused by an evolving one.

Here, we intend to consider as the non--linearity of the field evolution the 
\emph{Hall--effect} or \emph{Hall--drift}, as it occurs in the crust
\citep[see, e.g.,][]{SU97}, i.e., both ambipolar diffusion and any convective motion of
the neutron star matter is excluded.\\
The effect of the Hall--drift on the magnetic field evolution
of isolated neutron stars has been considered by a number of authors  
\citep[see, e.g.,][]{HUY90,GR92,M94,NK94,SU95,SU97,US95,US99,VCO00}.  
They discussed the redistribution of magnetic energy from an initially
large--scale (e.g., dipolar) field into small--scale
components via the Hall--term. Though the Hall--drift itself is a non--dissipative 
process, these changes in the field geometry 
may in principle accelerate the field decay. 
However, the results presented in \citet{NK94}, \citet{M94},
\citet{SU97}, and \citet{US99}
suggest the conclusion that the field decay is not modified drastically.

Goldreich \& Reisenegger \citet{GR92} developed the idea of the 
{\em Hall--cascade}, i.e., when starting with a large--scale magnetic field
small--scale field components are generated down to a 
scalelength $l_{\text{crit}}$, where the ohmic dissipation begins
to dominate the Hall--drift. 

In some of the above--mentioned investigations numerical instabilities are 
mentioned if either the field structure becomes too complex \citep{US95} 
or the initial field is too strong \citep{NK94,US99}.
Also, when considering the thermomagnetic field generation in the crust
of young neutron stars \citep{WG96}, where
small--scale modes are the first ones to be excited, 
numerical instabilities occurred exclusively caused by the Hall--drift.

Recently we have shown \citep{RG02} that 
all the observed instabilities are in their essence very likely not of
numerical origin but have physical reasons: A sufficiently strong and inhomogeneous
large--scale background field is unstable with respect to small--scale perturbations, which
rise very rapidly in comparison with the relevant ohmic decay time
of the background field.
This rapid transfer of energy from the background into small--scale field components proceeds \emph{not}
via a cascade but jump--like across wide spectral distances.
The unstable perturbations show 
small radial structures close to the neutron star's surface whereas their lateral structures are of
medium size.
Of course, this Hall--instability 
can act only during an episode of the field decay, which unavoidably
leads to a zero field. This episode, however, may have observable
consequences. 

In the following we will present the theoretical and observational hints for the
non--linear field decay in isolated neutron stars. After a 
description of the Hall--drift induced field instability we discuss its
effects on observable quantities of the above introduced SOINSs.

\section{Theoretical motivation}
\label{theomot}
Under which conditions becomes the evolution of the magnetic field non--linear?
The field evolution is determined by Maxwell's equations in quasi--stationary approximation
\begin{equation}
\hspace*{-5mm}\dot{\Bvec} = - c\,\curl \Evec\;,\quad
\dv \Bvec = 0\;,\quad \jvec =\frac{c}{4\pi}\curl{\Bvec}
\hspace*{-5mm}\label{eq:indeq}
\end{equation}
together with Ohm's law, which has in the absence of convective motions the form
\begin{equation}
\Evec = \hat{\sigma}^{-1}\jvec = \hat{R}\jvec~~.
\hspace*{-5mm}\label{eq:Ohmeq}
\end{equation}
In the presence of a magnetic field the electric 
conductivity, $\hat{\sigma}$, and the resistivity, $\hat{R}$, respectively, become
tensors whose components parallel and perpendicular to the magnetic field  
along with their Hall--component are represented by 
\begin{equation}
\hspace*{-1.8cm}\hat{\sigma} =\left(\begin{array}{lll} \phantom{-}\sigma_\perp& \sigma_\tH & 0\\
                                                    -\sigma_\tH&\sigma_\perp&0\\
                                                 \phantom{-}0&0&\sigma_\parallel \end{array}\right) =
\hat{R}^{-1} = \left(\begin{array}{lll} \phantom{-}R_\perp& R_\tH & 0\\
                                                     -R_\tH&R_\perp&0\\
                                              \phantom{-}0&0&R_\parallel
                     \end{array}\right)^{-1}\!\!\!\!\!\!,
\hspace*{-1.5cm}\label{tensor}
\end{equation}				
if $\Bvec$ goes along the $\vec{z}$--axis.

Let us consider the evolution of a magnetic 
field completely confined in the crust of the neutron star. There, the field
could have been generated either by  convection in the outermost layers of the
proto--neutron star or by a thermoelectric instability during the early period
of the neutron star's life, when the temperature gradient in the liquid crust is
extremely large \citep[see, e.g., ][]{ULY86,WG96}. A peculiarity of the transport processes in the crust is that
the components of the resistivity tensor parallel and perpendicular to the
magnetic field coincide \citep[see][]{YS91}, 
because the electrons move as a degenerate gas through the 
crystal lattice of the crust. In relaxation time approximation they found 
\begin{equation}
\hspace*{-5mm}\sigma_\parallel = \sigma_0\,,\quad  \sigma_\perp = \frac{\sigma_0}{1+(\omega_B\tau)^2} \,,\quad
     \sigma_\tH =  \omega_B\tau \, \sigma_\perp,
\hspace*{-5mm}\label{sigma}
\end{equation}
where $\omega_B = eB/m_e^{\ast}c$ is the Larmor 
frequency of the electrons, $ m_e^{\ast}$ their effective mass, and $\tau$
their relaxation time between two collisions with the phonons or impurities of
the crustal lattice. As a consequence of \Eqref{sigma} the parallel and perpendicular components
of the resistivity tensor just coincide 
\begin{equation}
\hspace*{-5mm} R_\parallel = R_\perp = \frac{\sigma_\perp}{\sigma_\perp^2+\sigma_\tH^2}
=\sigma_0^{-1}\,,\quad
     R_\tH =  \frac{\sigma_\tH}{\sigma_\perp^2+\sigma_\tH^2}.
\hspace*{-5mm}\label{R-sigma}
\end{equation}
The scalar electric conductivity $\sigma_0$ as well as the
relaxation time $\tau$ are complicated functions of the density 
$\rho(r)$, the temperature $T(r,t)$, the impurity concentration $Q$ and of the
chemical composition \citep[i.e., charge number $Z(r)$ and mass number $A(r)$; see, e.g.,][]{UY80}.

Ohm's law \eqref{eq:Ohmeq} together with \Eqref{tensor} allows of representing 
the electric field by use of
the unit vector of the magnetic field 
$\eBvec =\Bvec/|\Bvec|$ as
\begin{equation}
\hspace*{-9mm}\Evec = R_\parallel(\eBvec \cdot \jvec)\eBvec +
R_\perp(\eBvec\times\jvec)\times\eBvec + R_\tH(\jvec\times\eBvec),
\hspace*{-7mm}\label{E-general}
\end{equation}
which because of $ R_\perp = R_\parallel$ yields

\begin{equation}
\hspace*{-6mm}\Evec = R_\parallel\jvec + R_\tH(\jvec\times\eBvec) =
\sigma_0^{-1}\big(\jvec + \omega_B\tau(\jvec\times\eBvec)\big).
\hspace*{-5mm}\label{E-special}
\end{equation}
Taking now the $\curl$ of $\Evec$ and using Ampere's law we arrive at the 
\emph{Hall--induction equation} for the crystallized neutron star crust:
\begin{equation}
\hspace*{-1cm}\dot{\Bvec}= - \frac{c^2}{4\pi}\curl\bigg(\frac{1}{\sigma_0}
\big(\curl\Bvec+ \omega_B\tau\,(\,\curl\Bvec\;~~
\times\eBvec\,)\big)\!\bigg)
\hspace*{-5mm}\label{equ:Hall-Ind-Equ}
\end{equation}
which makes immediately clear that the field decay will proceed in a significantly
non--linear manner as soon as (and where) 
\begin{equation}
\frac{eB(r,t)}{m_e^{\ast}(r)c}\tau\big(\rho(r),T(r,t),Q,A(r),Z(r)\big) > 1\;,
\hspace*{-5mm}\label{equ:nonlincrit}
\end{equation}
i.e., the significance of the non--linearity 
depends not only on the field but on all the complex physical conditions in the crust.

In order to get an impression how the magnetization parameter $\omega_B\tau$ in
the crust of an isolated neutron star evolves we present here results obtained
by \citet{PGZ00} for a neutron star whose state of matter in
the core is described by a medium equation of state. The initial polar
surface field strength is assumed to be $10^{13}$ G, the density up to which 
the field is initially penetrating is $\rho_0 = 10^{14}$g cm$^{-3}$; for the 
impurity concentration $Q = 0.01$ is supposed and the chemical composition 
of the crust should be that of cold catalyzed matter.
\begin{figure}
\begin{center}
\epsfig{file=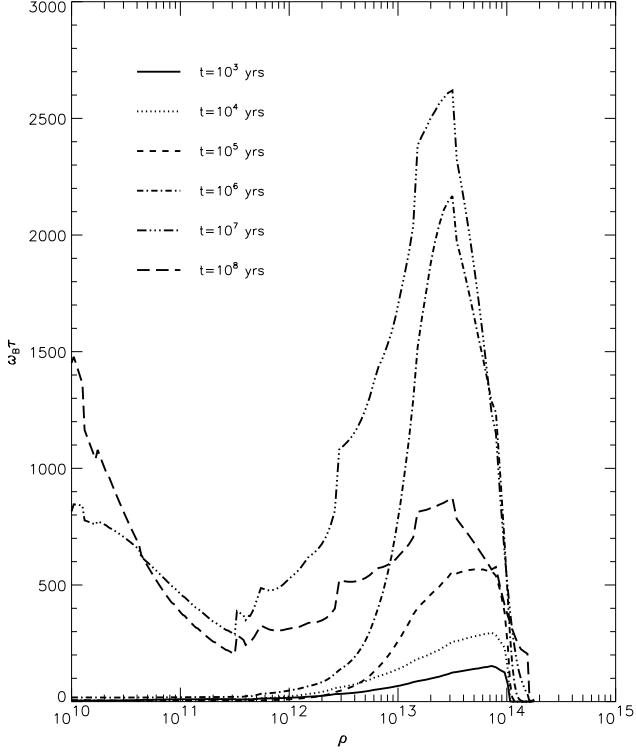,width=\linewidth,clip=} 
\end{center}
\caption{The magnetization parameter $\omega_B\tau$ as a function of the
density at different moments of neutron star's life.}
\label{omegatau}
\end{figure}
Since the field decay as presented by \citet{PGZ00} has been
calculated just by neglecting the non--linear effects, i.e., considering only the linear
ohmic decay, Fig.~\ref{omegatau} does not present the ``real'' evolution of
$\omega_B\tau$ because $\Bvec(r,t)$ certainly evolves differently under
the influence of a strong Hall--drift. Nevertheless, these results may serve
as a strong hint that the latter may play an important role for the
magnetic field decay in neutron stars. When, e.g., $\omega_B\tau$ becomes as large as
$1000$ as a consequence of the star's cooling down due to which the electron
relaxation time rises by orders of magnitude, a linear description of the field
decay is surely no longer justified.
                 
\section{Observational motivation}
\label{obsmot}
The best known observational quantities of SOINSs
are obtained when they appear as radiopulsars: their pulsation
period $P = 2\pi/\Omega$ ($\Omega$ being the related rotational velocity), its
time derivative $\dot{P}$ and, unfortunately only in a few cases, its second time derivative
$\ddot{P}$. 
The spin--down behaviour is to a high accuracy determined by the requirement, that
the loss of rotational energy ${E}_{\text{rot}} = I\Omega^2/2$
($I$ is the moment of inertia of the neutron star) is equal to the power of
the magneto--multipole radiation $\dot{E}_{\text{mmr}}$, emitted by the
rotating magnetized neutron star appearing as a radio pulsar:  
\begin{equation}
\dot{E}_{\text{rot}} = I \Omega \dot{\Omega} \approx -\dot{E}_{\text{mmr}}.
\hspace*{-5mm}\label{equ:spin-down}
\end{equation}
As already discussed above,
for pulsars older than $\approx 10^5$ years, the spin--down as a consequence of the
emission of gravitational waves can surely be neglected, too. The loss of 
angular momentum due to a
relativistic wind of charged particles (caused by the pulsar mechanism) shows the
same functional dependences on the neutron star radius $R$, the magnetic field
and the rotational velocity as 
the magneto--multipole radiation \citep{ST83}. 
Clearly, \Eqref{equ:spin-down} is not valid during glitches.

According to \citet{K91} the power of the 
magneto--multipole radiation is given by
\begin{equation} 
\hspace*{-7mm}\dot{E}_{\text{mmr}}=\frac{c}{8\pi} \sum_{l,m}
S_{lm}^{-2}\left[\frac{2^ll!}{(2l)!}\right]^2\!\!\left(\frac{mR\Omega}
{c}\right)^{2(l+1)}\!\!R^2\langle|\vec{B}_{lm}|\rangle^2,
\hspace*{-7mm}\label{equ:Krolik}
\end{equation}
where $ \langle|\vec{B}_{lm}|\rangle$ is the average of the multipole component
of the magnetic field, defined by its multipolarity $l$ and projection $m$,
across the spherical surface.
$S_{lm}$ is a constant reflecting the geometry of $\vec{B}_{lm}$ 
\citep[see Krolik,][]{K91}. Since $R$ and $I$ are fixed for 
SOINSs, we have with 
Eqs.~\ref{equ:spin-down} and \eqref{equ:Krolik} an unambiguous relation between the
observational quantities $P$ and $\dot{P}$ and the magnetic field. 
In other words, if we were able to calculate
all the multipole components of the magnetic field as functions of 
time and if these together with the measured  quantities
$P(t)$ and $\dot{P}(t)$ satisfied \Eqref{equ:spin-down}, we had understood the 
field decay. 
However, as a consequence of the sum occurring in \Eqref{equ:Krolik} it is
impossible to infer from $P(t)$ and $\dot{P}(t)$ all the 
$\langle|\vec{B}_{lm}|\rangle$ in an unambiguous way. Therefore, we consider here
only one term of the sum to be responsible for the spin--down and denote
the corresponding $\langle|\vec{B}_{lm}|\rangle$ simply by $B$. (Usually, the dipolar term,
$l=m=1$, is chosen.)
Then, by equating 
\Eqref{equ:spin-down} with \Eqref{equ:Krolik}, the change of the neutron star's
rotation can be written as 
  \begin{equation} 
\dot{\Omega} = - K \Omega^n  
 \hspace*{-5mm}\label{equ:omegadot}
\end{equation}
where the constant $K$ comprises the geometrical constant $S_{lm}$ as well 
as radius and moment of inertia and the magnetic field. 
The exponent $n=2l+1$ is
called the ``\emph{braking index}'', where $n = 3$ for a pure 
rotating dipole. The standard way to calculate the braking index is to take the time derivative
of \Eqref{equ:omegadot} resulting in
\begin{equation} 
 n = \frac{\ddot{\Omega}\Omega}{\dot{\Omega}^2} = 2 -
\frac{\ddot{P}P}{\dot{P}^2} \;.
\hspace*{-5mm}\label{equ:BIstandard}
\end{equation} 
However, $\ddot{\Omega}$ (or $\ddot{P}$) can be obtained with sufficient 
accuracy for a few young pulsars only.

\citet{JG99} suggested an
alternative way to calculate $n$ without employing 
$\ddot{P}$. By direct integration of \Eqref{equ:omegadot} over a time 
interval $T \gg P$
\begin{equation}  
\int_{\Omega_1}^{\Omega_2} \frac{\text{d}\Omega}{\Omega^n} = -K T
\hspace*{-5mm}\label{omegaint}
\end{equation}  
they calculate the braking index by
\begin{equation}
 n = 1 + \frac{\Omega_1\dot{\Omega}_2 -
\Omega_2\dot{\Omega}_1}{\dot{\Omega}_1\dot{\Omega}_2 T}\,,
\hspace*{-5mm}\label{BIalternative} 
\end{equation} 
i.e., knowing the rotational period and its time derivative of two
observations, separated by $T$, one is able to obtain the
braking index without needing $\ddot{P}$. For the Crab pulsar and B 1509--58
they found a good coincidence of the values of $n$ obtained with both methods.
Thus, there
exists a certain warranty that the application to other radio pulsars will 
yield reliable results, too. In their Table 1 
\citet{JG99} present the braking index $n$, calculated according to \Eqref{BIalternative}
along with $P$ and 
$\dot{P}$ for 20 pulsars. Six of them have
negative braking indices which possibly could be understood as due to the occurrence of glitches
or the continuation of a post--glitch recovery
during the observational periods.
As will become clear later, a {\em decay} of the NS magnetic field
can never be responsible for these negative values (whereas a field {\em growth} can).
The eight pulsars with the smallest error 
bars for $n$ (\object{B0540+23},
\object{B0611+22}, \object{B0656+14}, \object{B0740--28}, \object{B1915+13}, \object{B2002+31}, \object{B2148+52} and \object{B2334+61}) have
all positive braking indices, five of them with $n > 20$. Because of their 
similar braking indices \object{B0919+06}, \object{B1221-63} and \object{B1907+10}
may belong to that group too, but their error
bars are much larger. Two further pulsars, \object{B0154+61} and \object{B1356-60}, with $n>3$ have unacceptably
large error bars. 
By use of a recently available pulsar catalogue \citep{M01} we found two more SOINSs
(\object{B0727-18}, \object{B1822-09})
for which also earlier observations exist \citep{TML93}.
Their derived values of $n$ are the largest found by the above method, although the
error of $n$ for \object{B1822-09} is rather high (about 60\%).  
All pulsars mentioned have an
active age $\tau_a= P/2\dot{P}$ between $4.13\cdot 10^4$(\object{B2334+61}) and 
$1.7\cdot10^6$ (\object{B1907+10}) years, i.e., are just at the range of ages, 
the SOINSs should be. For a collection of the selected SOINSs' data see Table~\ref{table}.

A basic supposition for both methods to calculate the braking
index is that the coefficient $K$ in \Eqref{equ:omegadot}, by virtue of 
\Eqref{equ:Krolik} proportional to $B^2$, is 
really a constant. Now let us relax that assumption in so far as we will allow
for a varying field while the other quantities entering $K$ remain constant, 
in agreement with our
concept of SOINSs. Then we can rewrite 
\Eqref{equ:omegadot} into
\begin{equation}
\dot{P} = {\cal{K}} B^2(t)P^{-n+2}\,,
\hspace*{-5mm}\label{equ:spin-down-B(t)} 
\end{equation} 
where for a standard neutron star ($R=10^6$cm, $I=10^{45}$gcm$^2$) the
constant ${\cal{K}} \approx 10^{-39}$ cm s$^3$g$^{-1}$ if $n=3$. 
We will nevertheless assume in the following that the rotational evolution is during a period $T$
still quite well described by a power--law similar to \Eqref{equ:omegadot}, where, of course, 
the quantities $K$ and $n$ have to be replaced by modified ones:
\begin{equation}
\dot{\Omega} = - {\hat{K}} \Omega^{\hat{n}}\,.  
\hspace*{-5mm}\label{equ:omegadot-hat}
\end{equation}
The description of the rotational behaviour by a power law should be 
sufficiently precise as long as the time $T$ (see \Eqref{BIalternative}) 
is small in comparison with the characteristic time of the magnetic field
evolution. Of course, a final justification for that assumption can only be 
obtained for each individual pulsar by considering its actual time series of
$P$ and $\dot P$.
Re--evaluating now 
\Eqref{equ:BIstandard} with $n$ replaced by $\hat{n}$ we find with 
$\ddot{P}={\cal{K}}BP^{-n+1}\left(2\dot{B}P\,\,-\,\,(n-2)B\dot{P}\right)$ 
the correction of $n$ to the ``observed'' 
${\hat{n}}$ to be proportional to the time derivative of the magnetic field:
\begin{equation}
{\hat{n}}=n-2\frac{\dot{B} P}{B \dot P} = n - 2\,\sign(\dot B\dot P)\frac{\tau_P}{\tau_B} \,. 
\hspace*{-5mm}\label{equ:hatn} 
\end{equation}
Here we introduced $\tau_P$ and $\tau_B$ as characteristic times of the  
rotational period and the magnetic field, respectively, but we discuss this relation
for a spin--down situation only, that is, $\dot P>0$ 
How should this equation be interpreted? In case there is no magnetic
field change, that is $\dot{B} = 0$, pulsars with $n < 3$ are not 
understandable in the framework of the magneto--multipole radiation model, 
because the neutron star will certainly not be a rotating monopole 
(with small higher multipole contributions). 
In case we assume that the spin--down is completely due to
magneto--\emph{dipole} radiation, an assumption perhaps not completely wrong 
for SOINSs, we have $n = 3$ and 
${\hat{n}} < 3$ indicates an increasing field while ${\hat{n}} > 3$ is the 
signature of a decaying one. 
The interpretation of ${\hat{n}} < 3$ as caused by an
increasing field is supported by the fact, that such braking indices are 
observed especially
for very young pulsars \citep{LPGC96}, 
in which both rediffusion of a field 
submerged during the supernova fallback \citep{GPZ99} and/or thermoelectric field
generation may still proceed. Alternatively, deviations of ${\hat{n}}$ 
from its classical value $3$, even  
values of either sign up to $10^6$, can be understood 
as due to internal frictional instabilities occurring between the crust
and the superfluid, almost independently on the evolution of the NS magnetic field 
\citep{SH95,SM95,K01}.
However, this explanation applies, except for strong vortex pinning, only for older 
(age $>$ 1.8 $10^7$ yrs) NSs. Thus the assumption of a rapid field decay could just explain 
large observed values of $\hat{n}$ for younger objects, as the SOINSs listed in Table~\ref{table} are.   

Also processes in the magnetosphere may cause values  
of ${\hat{n}}$ different from $3$, but these effects are considered to cause
$\hat{n}<3$ in young pulsars \citep[see, e.g.,][]{M97,CM98}.   

In this context it should be mentioned that the conclusion of \citet{RF01}
a decaying magnetic field causing a braking index greater than three was ``in disagreement with
observations" is obviously wrong. Along with those for which \citet{JG99} calculated
the braking index it has been determined for only five very young pulsars employing $\ddot{P}$.
In all there exists only a single case (J1119-6127) with almost exactly
$n=3$. Thus, the assumption of a varying (either increasing or decreasing) magnetic field is in
general the more appropriate one.

\begin{table}
\caption{\label{table} SOINSs supposed to be in the phase of non--linear magnetic field decay.
Rotational period $P$ and its time derivative $\dot{P}$ are mainly taken from Table 1 of \citet{JG99},
$\hat{n}$ is the rounded mean value of $n$ given there. The data for B0727-18 and B1822-09 come from
the pulsar catalogues \citet{TML93} and \citet{M01}.
The active age, polar dipole field
strength and field decay rate are calculated as described in Sect.~\ref{obsmot} }

\begin{center}
\begin{tabular}{@{}>{$}c<{$}c@{\hspace{3mm}}c@{\hspace{3mm}}c@{\hspace{3mm}}c@{\hspace{1mm}}c@{\hspace{1mm}}c@{\hspace{0cm}}}
\hline
\hline
\text{PSR}(B) & $P$   & $\dot{P}  $   & $\hat{n}$ & $\tau_{\text{a}}$     & $B_{\text{d}}$ & $\dot{B}_{\text{d}}$       \\*[.5mm]
\hline
           &   [s] & $ [10^{-14}]$ &           & [$10^5$ yrs] & [$10^{12}$G]   & [$10^8$G/yr]               \\
\hline
0540\!+\!23   & 0.246   & 1.531    & 12        & 2.55         & 3.93           & 0.35                       \\
0611\!+\!22   & 0.334   & 5.921    & 20        & 0.89         & 9.00           & 4.28                       \\
0656\!+\!14   & 0.385   & 5.507    & 15        & 1.11         & 9.32           & 2.52                       \\
0727\!-\!18   & 0.510   & 19.01    & 319       & 4.26         & 6.30	       &11.66			    \\
0740\!-\!28   & 0.167   & 1.687    & 26        & 1.57         & 3.40           & 1.25                       \\
0919\!+\!06   & 0.431   & 1.375    & 29        & 4.97         & 4.93           & 0.65                       \\
1221\!-\!63   & 0.216   & 0.493    & 19        & 6.94         & 2.09           & 0.12                       \\
1822\!-\!09   & 0.769   & 52.43    & 142       & 2.33         & 12.8	       &19.20			    \\
1907\!+\!10   & 0.284   & 0.265    & 24        & 17.6         & 1.75           & 0.05                       \\
1915\!+\!13   & 0.195   & 0.724    & 36        & 4.28         & 2.40           & 0.46                       \\
2002\!+\!31   & 2.128   & 7.562    & 23        & 4.46         & 25.7           & 2.88                       \\
2148\!+\!52   & 0.332   & 1.005    & 50        & 5.24         & 3.70           & 0.83                       \\
2334\!+\!61   & 0.495   & 19.02    & 9         & 0.41         & 19.6           & 7.13                       \\                              
\hline
\end{tabular}
\end{center}
\end{table}

Here we intend to investigate, whether the observed 
relatively large values of ${\hat{n}}$ may indicate a \emph{non--linear field decay}. 
For that
purpose let us consider as a typical example \object{B0740-28}, found in Table 1 of
\citet{JG99}, with ${\hat{n}} \approx 26$. From
its observed $P = 0.167$s, $\dot{P} = 1.68\cdot 10^{-14}$ 
we can infer a dipolar surface magnetic field strength $B_{\text{d}}=6.4\cdot
10^{19}\sqrt{P{\dot{P}}} \approx 3.4\cdot 10^{12}$ G which is in the range of values typical for
such ``middle aged'' radio pulsars. Using the standard formula
for the active age $\tau_a = P/2\dot{P}$ we find 
$\tau_a\approx 1.6\cdot 10^5$ years, but this relies on the assumption that
the rotational evolution follows all the time the power law \eqref{equ:omegadot-hat} 
with $\hat{n}=3$. Of course, this value of $\tau_a$ is at best an upper limit for the star's age, since we just argue
in favour of an $\hat{n}>3$ during some (unknown) period of the rotational evolution
still lasting. On the other hand a much shorter active age is unlike because none but
one of the mentioned pulsars is reported to be associated with a supernova remnant.
The only exception, \object{B2334+61}, with $\tau_a=4.1\cdot 10^4$ yrs, is associated with the
fairly old \object{SNR G114.3+0.3} \citet{BBT96}; thus the active age seems to be a sound estimate of the real
age. 

If we exclude the possibility of magnetic field decay we have for our example (\object{B0740-28})  
$\,\hat{n}\!=\! n \!\approx\! 26$, i.e., the neutron star's magnetic field structure must
be that of a $2^{12}$--pole because of $l \approx 12$. As a consequence of \Eqref{equ:Krolik} this implies, that the
surface magnetic field strength of such a field must be many orders of magnitudes stronger
than that of a dipolar field in order to
explain the observed spin--down. Therefore, the
assumption of a constant magnetic field seems not to be very likely. If we
instead assume that the spin--down is caused by magneto--dipole
radiation, i.e. $n=3$, we can easily estimate the rate of the magnetic field decay.
Rearranging \Eqref{equ:hatn} and inserting the observed or inferred values for
$P, \dot{P}$ and $B_{\text{d}}$ we find
\begin{equation}
\hspace*{-5mm}\dot{B}_{\text{d}}=(n-\hat{n})\frac{B_{\text{d}}\dot{P}}{2P} \approx 4\text{G/s}\approx
1.25\cdot 10^8\text{G/yr}\,.  
\hspace*{-5mm}\label{equ:spin-down-Bdot} 
\end{equation}
This rate of the dipolar--field decay should now be compared with
the typical ohmic decay rate of a crustal  magnetic field. 
The ohmic decay time of a crust of 
thickness $L=10^5\text{cm}$ and an averaged electric conductivity 
$\sigma_0=5\cdot10^{26}\text{s}^{-1}$ is
\begin{equation}
 \tau_{\text{ohm}}=\frac{4\pi\sigma_0
L^2}{c^2} \approx 2.2\cdot 10^9\text{yrs},
\hspace*{-5mm}\label{equ:ohmic-decay} 
\end{equation}
providing an upper bound for the real field decay times.
By use of the above estimated 
${B}_{\text{d}} = 3.4\cdot 10^{12}$ G we find
\begin{equation}
\hspace*{-5mm}\dot{B}_{\text{d}}= 1.25\cdot 10^8\text{G/yr} \gg
\frac{{B}_{\text{d}}}{\tau_{\text{ohm}}} \approx 1.5\cdot 10^3\text{G/yr}. 
\hspace*{-5mm}\label{equ:comparison} 
\end{equation}
This comparison is robust, i.e., even if we somewhat overestimated $L$
or $\sigma_0$ (which, because of the
unknown impurity concentration, could indeed also be underestimated),
the magnetic field decay   
proceeds obviously much faster than would be predicted for a (linear) ohmic decay
\citep[see, e.g.,][]{TK01}.
What has been demonstrated here for a single example proves to be valid for the complete 
selection of SOINSs specified above, see Table~\ref{table}. 

As demonstrated in Sect.~\ref{theomot} the only possible modification of the ohmic decay    
in the context of our model is due to the Hall--drift.
{\em That is, any significant acceleration of the magnetic field decay is explainable
then and only then, if the Hall--drift can be shown to cause time scales 
of the field evolution significantly smaller then the ohmic one.}
\begin{figure}
\begin{center}
\epsfig{file=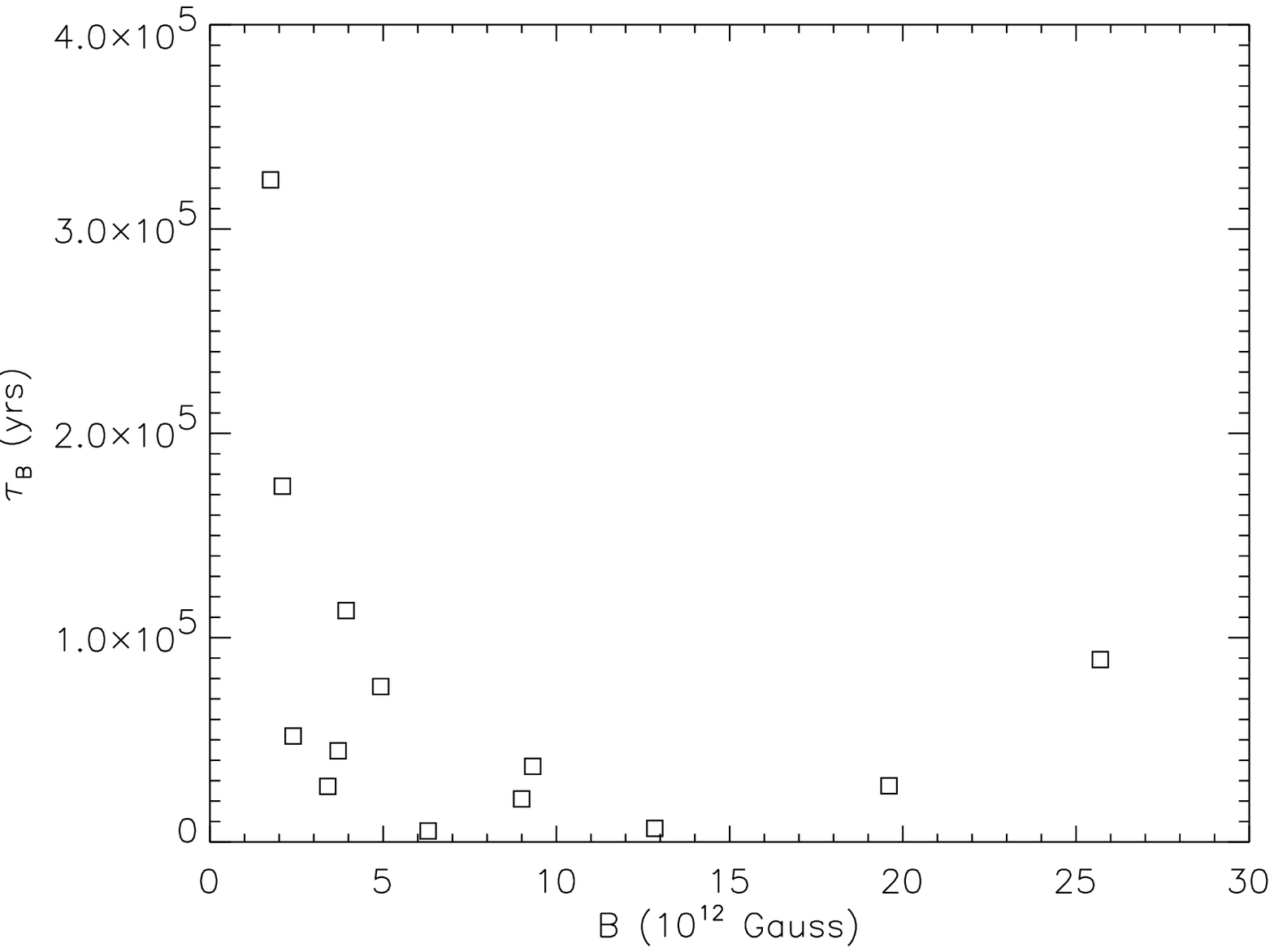,width=\linewidth, clip=}

\epsfig{file=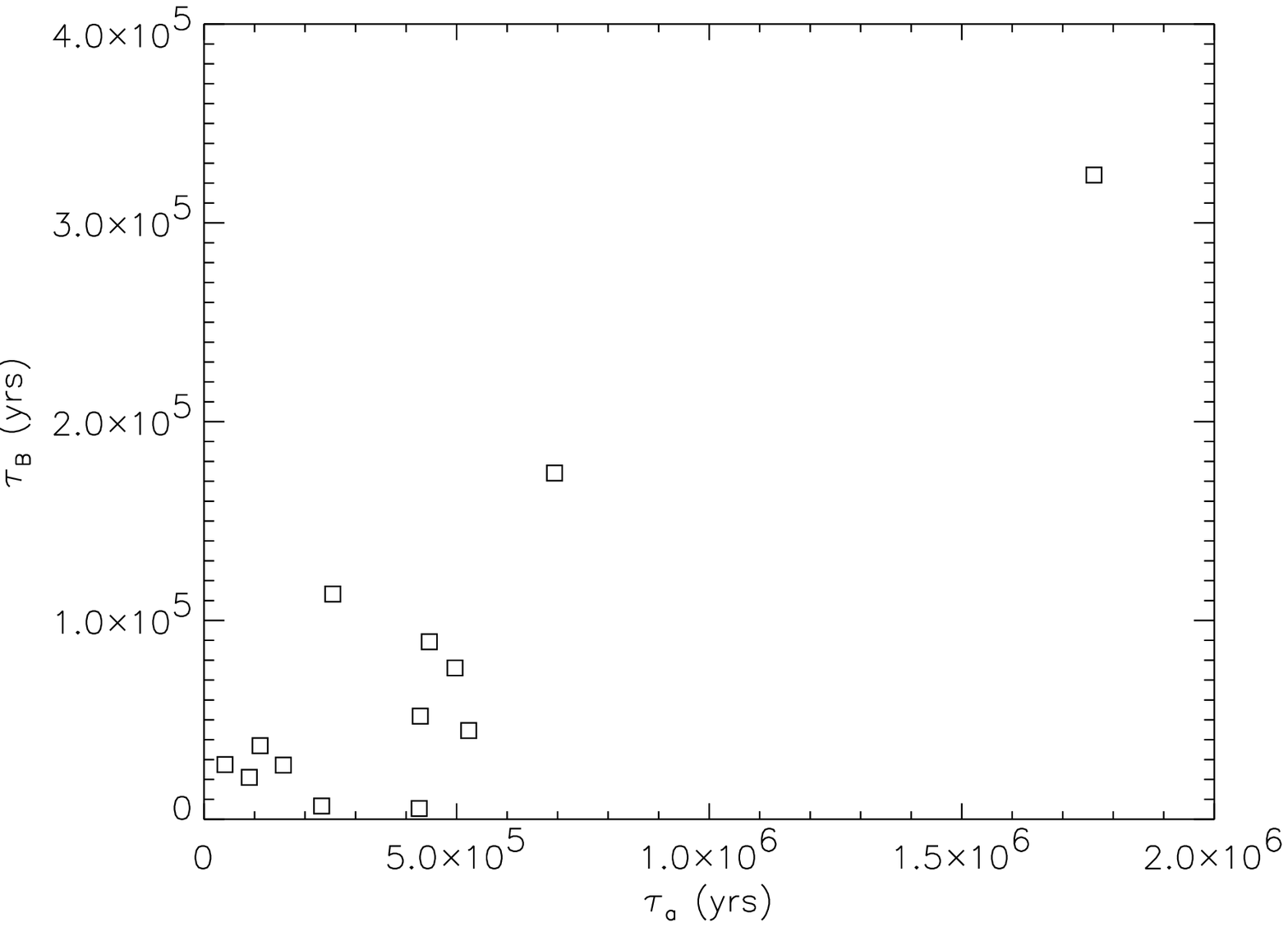,width=\linewidth, clip=}
\end{center}
\caption{\label{figkorr} Characteristic time of the magnetic field, $\tau_B$, versus the magnetic field
$B$ itself (upper panel) and versus the active age $\tau_a$ (lower panel) for the SOINSs given in Table~\ref{table}.} 
\end{figure}       
In order to visualize the clear signs of non--linear field decay for the selected SOINSs we present in the upper panel of Fig.~\ref{figkorr}
the correlation of the characteristic time of the magnetic field $\tau_B=|B_{\text{d}}/\dot{B_{\text{d}}}|$ with
the magnetic field $B_{\text{d}}$ itself. Bearing in mind, that thirteen objects
form only a very small statistical ensemble, there nevertheless seems to exist a pronounced trend
indicating non--linearity of the field decay: Evidently, $\tau_B$ falls significantly
with increasing $B_{\text{d}}$ which is compatible with the idea that the stronger the field the faster proceeds the
redistribution of energy into smaller modes by the Hall--drift and, consequently, the faster the energy
is drained from the dipolar field mode. Note, that even the two pulsars (\object{B0727-18}, \object{B1822-09}) with
apparently extraordinary high values of $\dot{B_{\text{d}}}$ fit well into this trend.

In addition, we show in the lower panel of Fig.~\ref{figkorr} the correlation of $\tau_B$ with
the active age $\tau_a$ suggesting the conclusion that the acceleration of the field decay by the Hall--drift 
acts only temporarily. This makes sense because that effect of course destroys its own basis, namely
the dominance of the Hall--drift over the ohmic decay.      
Certainly it is to premature to take the link between the field decay and the
observations as a proven fact. However, the apparent trend gives enough reason
to pursue this idea.

In the following chapter we will present an instability which proceeds on
time--scales  much smaller than the ohmic scale.

\section{A Hall--drift induced magnetic field instability}

\subsection{Hints for the occurrence of an instability}
Now, let us return to \Eqref{equ:Hall-Ind-Equ} which together with $\dv \Bvec
= 0$ and appropriate boundary conditions determines the magnetic field evolution
in the crust of isolated neutron stars.
It is generally accepted that the non--linear Hall--term causes a
re--distribution of magnetic energy, when it is at a certain ``initial'' state
stored in the mode characterized by the largest spatial scale, say a dipolar
field mode. \citet{GR92} coined the term 
\emph{Hall--cascade} for that process by stressing an analogy to the vorticity 
equation for an incompressible fluid. As they noted, this analogy is incomplete 
at least as far as one $\curl$ operator has to be replaced by its inverse, 
which makes the conclusions about the Kolmogoroff spectrum questionable. 

Attempts performed by the present authors to solve the non--linear \Eqref{equ:Hall-Ind-Equ}
in a full sphere assuming a constant density profile 
reproduced the results of \citet{SU97} for moderate values of
the magnetization parameter $\omega_B\tau$. Starting with a dipolar field modes
of smaller scales were
excited and helicoidal waves were observed. But, as soon as
$\omega_B\tau$ exceeded a certain value, typically about
$100$, the spectral resolution, attainable by the numerical code
was no longer sufficient to ensure convergence because the amplitudes of
small--scale modes at the tail of the resolvable spectrum started to grow very quickly.
This motivated us
to search for possible physical reasons for that apparent
numerical instability.

For simplicity we assume the conductive properties of 
the matter to be constant in space and time, that is, we assume constant
$\sigma_0$ and $\tau_e/m_e^*$. Thus, \Eqref{equ:Hall-Ind-Equ}
can be rewritten in dimensionless variables such that
it no longer contains any parameter and the evolution of the magnetic field is solely
determined by its initial configuration $\Bvec(\xvec,0)$.
For that purpose we normalize the
spatial coordinates by a characteristic length $L$ of the model
(e.g., the neutron star's radius or the thickness of its
crust), the time by the ohmic decay time $4\pi\sigma_0 L^2/c^2$ and the
magnetic field by $B_N=m_e^* c/ e\tau_e$. The
governing equations in these dimensionless variables read 
\begin{equation}
\hspace*{-5mm}\dot{\Bvec} = \Delta\Bvec - \curl (\,\curl\Bvec
\times\Bvec\,)\,\;,\;\;\dv \Bvec = 0\;, 
\hspace{-5mm}\label{equ:indeqdimless}
\end{equation}
where the differential operations have to be performed with respect to the
now dimensionless spatial and time coordinates
$x,y,z$ and $\tau$, respectively. 

Let us first have a look onto energetics.
Using standard arguments one can immediately state that in the absence
of currents at infinity the total energy of any
solution of \Eqref{equ:indeqdimless} is bound to decrease monotonically to zero since the
Hall--term $\curl\Bvec\times\Bvec$ is unable to deliver energy
(nor to consume it). That is, all these solutions are asymptotically
stable in a strict mathematical sense following Lyapunov's definition of
stability. However, this applies also to a lot of slowly time varying processes (background)
for which nevertheless
the transient occurrence of instabilities is a generally accepted idea (e.g.,
the temporary convection in a cooling liquid). Therefore, it seems reasonable to
adopt as a criterion
of instability the following: Define a steady reference (equilibrium) state
in the spirit of the usual stability analysis to be a snapshot of the background process. If
the analysis detects growing, that is, unstable perturbations with a time increment
large enough in comparison with the time decrement of the background process under
consideration the latter may be accounted unstable in a short enough period of time
around the instant of the snapshot.   

Of course, when adopting this concept a necessary condition for the occurrence of an instability
is as well the existence of terms in the linearized equation which are potentially able to deliver energy
to the perturbations.  
Linearization of \Eqref{equ:indeqdimless} about the {\em background field} $\Bvec_0$ yields
\begin{equation}
\hspace*{-5mm}\begin{aligned}
&\dot{\bvec} = \Delta\bvec - \curl (\,\curl\Bvec_0
\times\bvec\ + \curl\bvec\times\Bvec_0\,)\\ &\dv \bvec = 0 
\end{aligned}
\hspace*{-5mm}\label{equ:indeqdimlesslin}
\end{equation}
describing the behavior of small perturbations $\bvec$ of the reference
state $\Bvec_0$.
Indeed, here along with the energy--conserving term $\curl\bvec\times\Bvec_0$ 
now as a second Hall--term $\curl\Bvec_0\times\bvec$ occurs
which may well deliver or consume energy (from/to $\bvec$) since in 
general the integral 
$\int_V (\curl\Bvec_0\times\bvec)\cdot \curl\bvec\, dV$ will not vanish.
Of course, this is due to the fact that the linearized equation describes 
the behavior of only a part of the total magnetic field. Actually, 
perturbations
may grow only on expense of the energy stored in the background field.

Considering \Eqref{equ:indeqdimlesslin}, we can determine a scale below
which the
ohmic dissipation dominates the Hall--drift. Estimating $|\curl\Bvec_0|$
and $|\curl\bvec\,|$ by $\bar B_0$ and $\bar b / l$, respectively,
we find the critical scale of $\bvec$ to be $l_{\text{crit}} \lesssim 1/\bar B_0$, 
which is de--normalized $l_{\text{crit}} \lesssim L/(\omega_{{\bar B}_0}\tau_e)$, 
identical with the expression derived by \citet{GR92} considering the 
Hall--cascade in analogy with the turbulent flow of an incompressible fluid.

\subsection{Analogy with the $\Omvec - (\omvec \times
\jvec)$--dynamo}

A strong support for the idea of a Hall--drift induced magnetic
field instability is provided by the analogy of the linearized Hall--induction
equation \eqref{equ:indeqdimlesslin} to the mean--field induction equation including the
so--called $(\omvec \times \jvec)$--effect found by \citet{R69}:
\begin{equation}
 \hspace*{-5mm}\dot{\bar{\Bvec}} = \lapl\bar{\Bvec} - \curl\, (\,-\uvec_{\Omega}\times\bar{\Bvec} + \curl\bar{\Bvec}\times \omvec\,)\,.
\hspace{-5mm}\label{equ:raedler}
\end{equation}
Here, $\bar{\Bvec}$ is the averaged magnetic field, $\uvec_\Omega$ describes a (large--scale) differential rotation and $\omvec$ is
a turbulence--related coefficient,
which reflects the offset of the isotropy of the turbulence by Coriolis
forces.
\Eqref{equ:raedler} has been proved to allow for magnetic instabilities \citep[i.e., dynamos; see][]{R69}.

Comparing \Eqref{equ:indeqdimlesslin} with \Eqref{equ:raedler} when letting correspond $\bvec$
to $\bar{\Bvec}$ it becomes clear, that $\curl\Bvec_0$ (then corresponding to $-\uvec_{\Omega}$)
must not be a constant, since $-\uvec_{\Omega}$ is a shear velocity. This implies that not all
second derivatives of $\Bvec_0$ with respect to the spatial co--ordinates may vanish.   
It should be noted that the analogy lacks completeness only insofar
that the dynamo acts even with a constant $\vec{\omega}$.
Note that $\vec{\omega}$ enters the term which is energy conserving.

\subsection{Mathematical treatment}
Let us now specify the geometry of our model and 
the background field. We consider a slab which is infinitely extended
both into the $x$-- and $y$--directions but has a finite thickness $2L$ in
$z$--direction.
\begin{figure}
\setlength{\unitlength}{1.mm}
\hspace*{-1cm}\begin{picture}(90,25)
\thicklines
\put(10,1.2){\colorbox{hgrau}{\makebox(68,13.75){}}}
\put(43,-3){{\small $z=-1$}}
\put(10,0){\line(1,0){70}}
\put(10,16){\line(1,0){70}}
\put(43,18){{\small $z=+1$}}
\quadfeld(40,2.7)(0,1){l}\put(58,8){$\Bvec_0$}
\dreibein(85,8){10}{5}{2}{8}{1}
\end{picture}

\vspace{2mm}
\caption{\label{geometrie} Sketch of the model and background field geometry.
In $z<-1$ a perfect conductor, in $z>1$ vacuum is assumed.}
\end{figure}
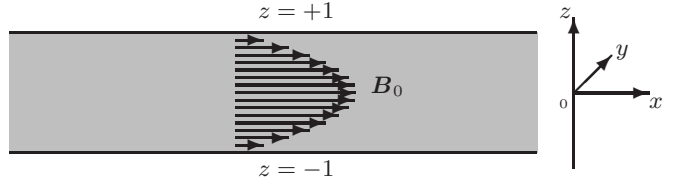

The background field $\Bvec_0$ is assumed to be everywhere
parallel to the surface of the slab pointing, say, in $x$--direction and to
depend on the $z$--coordinate only, i.e., $\Bvec_0 = f(z) \exvec$ ($\exvec$ 
being the unit vector in $x$--direction). 
Note, that by this choice $\curl\Bvec_0\times \Bvec_0$
represents a gradient. Thus, the unperturbed evolution of the
background field is not at all affected by the Hall--drift; in the
absence of an e.m.f. it would decay purely ohmically.

Further on we decompose a perturbation $\bvec$ into its
poloidal and a toroidal components, $\bvec =  \bvec_\tP+\bvec_\tT$, which can
be  represented by scalar
functions $S$ and $T$, respectively, by virtue of the definitions
\begin{equation}
\hspace*{-7mm}\bvec_\tP = -\curl\,(\,\ezvec \times \nabla S)~~,\; 
\bvec_\tT = -\ezvec \times \nabla T\;, 
\hspace*{-1cm}\label{poltorST}
\end{equation}
ensuring $\dv\bvec=0$ for arbitrary $S,T$.

For the sake of simplicity we will confine ourselves to the
study of plane wave solutions with respect to the $x$-- and $y$--directions,
thus making the ansatz
\begin{equation}
\left\{\begin{aligned}
&S\\
&T
\end{aligned}\,\right\}(\xvec,\tau) = 
\left\{\begin{aligned}
&s\\
&t
\end{aligned}\,\right\} (z) \exp{(\ti\tilde{\kvec}\tilde{\xvec} + p\tau)}~~,
\label{transf}
\end{equation}
where $\tilde{\kvec} = (k_x,  k_y)\neq\zervec$, $\tilde{\xvec} = (x,y)$
and $p$ is a complex time increment.
This ansatz guarantees as well the uniqueness of the poloidal--toroidal decomposition
since from $\undertilde{\lapl}\,(S,T)=0$ it follows $(S,T\,)=0$ with
$\undertilde{\lapl}$ being the two--dimensional lateral Laplacian \cite[see][]{BR88}.
With \Eqref{transf} we obtain from \Eqref{equ:indeqdimlesslin} two coupled ordinary
differential equations
\begin{equation}
\hspace*{-1.4cm}\begin{alignedat}{4}
&pt -t'' &&+ \tilde{k}^2t &&= &&\ti k_x f (s'' -\tilde{k}^2s) - \ti k_x f''s\\ 
&ps -s'' &&+ \tilde{k}^2s - \ti k_y  f's &&=  - &&\ti k_x f t \;,
\end{alignedat}
\hspace*{-9mm}\label{poltoreq}
\end{equation}
where dashes denote derivatives with respect to $z$. 
Together with appropriate boundary conditions Eqs. \eqref{poltoreq} define an
eigenvalue problem with respect to $p$.

Having in mind that the neutron star's crust is
neighboured upon the superconducting core on the one and a region with very low conductivity
on the other side we choose the following simplified boundary conditions:  
Outside the star we assume vacuum, that is, $\curl \Bvec = \zervec$ above the slab.
Because of the finite conductivity inside the slab we require continuity of all components
of $\Bvec$ across this boundary.
Considering the crust--core boundary an electric field must be prevented from
penetrating into the perfectly conducting region below the slab, $z < -1$, that is, the normal magnetic and
tangential electric field components must vanish. In terms of the scalars $s$ and $t$ this means
$[s]=\left[s'\right]=t=0$
for the vacuum boundary and
$s=t'=0$
for the perfect conductor boundary where $[.]$ denotes the jump across a
boundary. Note, that for $t'=0$ to be valid the vanishing of $\Bvec_0$ at the
perfect conductor boundary is required.
Making
use of the vacuum solution for the upper halfspace, $z\ge 1$, vanishing at infinity, 
the vacuum boundary condition for $s$ 
can be expressed as $s' = - \tilde k s$ at $z= 1$, with 
$\tilde k = |\tilde{\kvec}|$.

Finally we specify the background field profile: Guided by the conditions under which
the above discussed $\Omvec - (\omvec \times\jvec)$--dynamo may work we conclude that
$f(z)$ has to be at least quadratic in $z$. 
Taking into account the boundary conditions we choose
$f(z)=B_0(1+z)(1-z^2)$. From now, the coefficient $B_0$ is the only free parameter of the
eigenvalue problem.

\subsection{Numerical results}
\begin{figure}
\hspace*{-.0cm}\epsfig{file=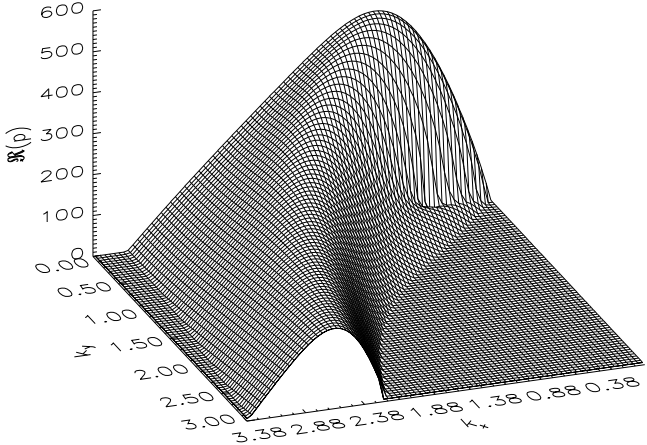, width=1.04\linewidth, clip=}
\caption{Growth rate as a function of $k_x$ and $k_y$ for
$B_0=1000$. Negative values were set to zero.}
\label{fig1}
\end{figure}

For certain ranges of the wave numbers ${k_x,k_y}$ and values of $B_0 \gtrsim 3$ we
found eigenvalues $p$ with a positive real part, i.e., exponentially growing
perturbations.
The dependence of
the growth rate $\Re(p)$ on the wave numbers for $B_0=1000$ is shown in Fig.~\ref{fig1}.

\begin{figure}[t]

\vspace*{-1.1cm}
\hspace*{-3.5mm}\epsfig{file=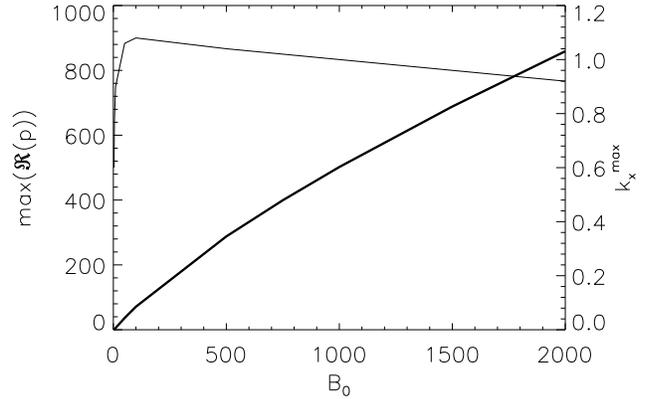,width=1.05\linewidth,clip=}

\vspace*{-4mm}
\caption{Growth rate and wave number $k_x^\tmax$ of the fastest growing mode as
functions of $B_0$.
Thick and thin lines correspond to growth rates and $k_x^\tmax$, respectively.}
\label{fig2}
\end{figure}

Figure \ref{fig2} shows the dependence of growth rate and wave number $k_x^\tmax$
of the fastest growing mode on $B_0$.
An interesting feature is, that the maximum growth rates occur
for all $B_0$ considered at $k_y=0$.
Of course this asymmetry is due to the choice of the
background field: once it was chosen parallel to the $y$--direction the maximum
growth rates would occur at
$k_x=0$. Note that the most unstable eigenmodes are always non--oscillatory, though
oscillating unstable ones exist.

Evidently, for each $B_0$ above a certain threshold (say, $B_0=50$) there exists
a range of wave numbers within which the obtained growth rates are in agreement with
the constraint, formulated above: In comparison with the background field decay
the growth of the most unstable perturbations is a fast process;
thus we may consider it as ``episodically unstable''.

With respect to the asymptotic behavior $\sigma_0\!\rightarrow\!\infty$ for a fixed
(unnormalized) background field one has to note
that the time increment $p$ is normalized on the ohmic decay rate ($\propto \sigma_0^{-1}$).
From Fig.~\ref{fig2} it can be inferred $\Re(p)\propto B_0^q\,,\;q < 1$ for $B_0 \ge 100$,
which means that in the limit of negligible
dissipation the growth rate in physical units tends to zero.    

\begin{figure}[h]
\hspace*{-0.5mm}\epsfig{file=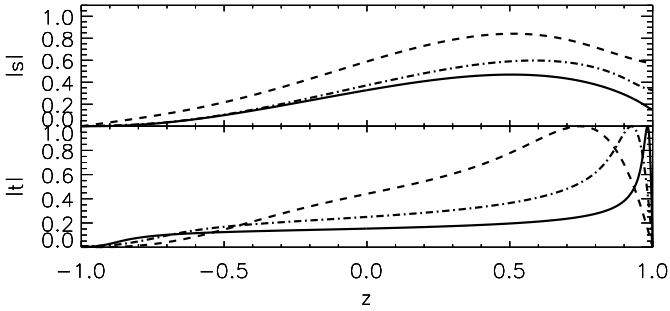, width=1.017\linewidth,clip=}\\*[-1mm]

\caption{Moduli of $(s,t\,)(z)$ of the fastest growing mode.
Solid, dash--dotted and dashed lines refer to $B_0=2000$, $B_0=100$ and
$B_0=10$, respectively.}
\label{fig5}
\end{figure}

\begin{figure}[h]
\begin{center}
\hspace*{-5mm}\epsfig{file=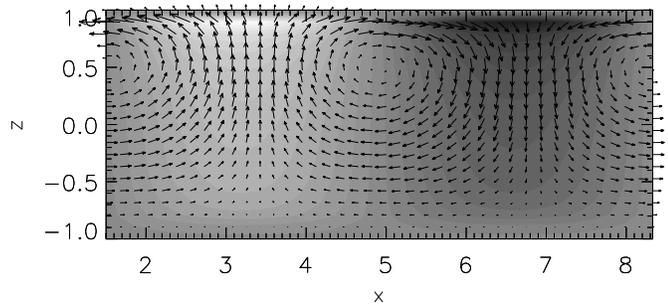, width=1.1\linewidth,clip=}
\end{center}

\vspace{-0.5cm}
\caption{Field structure of the fastest growing mode for $B_0=2000$.
Arrows: $b_{x,z}$, grey shading: value of $b_y$, dark -- into, light -- out
of the plane.}
\label{fig6}
\end{figure}

Figure \ref{fig5} shows the eigensolutions $(s,t\,)(z)$ of the fastest growing
mode for three different values of $B_0$.
One can observe that with increasing
$B_0$ the toroidal field
becomes more and more small--scale and concentrated towards the vacuum
boundary. In contrast, the corresponding poloidal field remains large--scale.
The magnetic field
structure of the fastest growing mode for $B_0=2000$ is shown
in Fig.~\ref{fig6}. Assuming a ratio of crust thickness and neutron star radius
$R$ equal to 0.1 the shown lateral period of that mode fits $\approx 20$ times
into the circumference of the star.
  

\section{Discussion and Conclusions}
Clearly, any assignment of the results gained by help of a very simplified
model to real neutron stars has to be done with great care. Even when 
accepting the plane layer as a reasonable approximation of the crust
one has to concede that the very specific profile of $\Bvec_0$ assumed above
may only exemplify the crustal field's structure. 

An acceptable approximation of the radial
profile of a dipolar crustal field as given, e.g., in \citet{PGZ00} will in
general have to allow for non--zero values at the boundaries which cause
a more complicated perfect conductor boundary condition. Moreover, the strong 
dependence of the conductive properties on the radial co--ordinate should anyway be
taken into account.

Recently, \citet{HR02} solved the non--linear \Eqref{equ:indeqdimless} for
the first time in a spherical shell geometry, again with the transport coefficients
constant in space and time, and applying the
vacuum  and the perfect conductor  boundary conditions at the surface and
at the crust bottom, respectively. This is the best
model for the non--linear magnetic field evolution in the crust of neutron stars
available so far.
They made different assumptions about the large--scale initial field
(which corresponds in some sense to our background field) and
report, as all the formerly mentioned authors too, about insurmountable
numerical problems when the initial magnetization parameter, very roughly to be identified 
with $B_0$, exceeds $\approx 200$.
Although their investigations can not reveal the existence and the
character of the instability presented above
(too short integration period in comparison with the growth
time of the unstable modes for the maximum initial field considered),
they found at least two
features which support our results:
For the largest feasible initial field their
Legendre spectra are no longer convergent and indicate a local minimum at
$l\approx 60$, a hint that small--scale field structures are fed in a
non--cascade--like, possibly unstable way.
Moreover, their findings confirm the importance of the background field's
profile curvature:
For an initial (poloidal) field nearly linear in $r$ 
the excitation of small--scale structures and the acceleration of the decay
were insignificant, in contrast to an initial
(toroidal) field with a nearly quadratic $r$--profile.
 
To get an impression of possible consequences for the evolution of neutron
stars
we now simply assume, that the real $\Bvec$--profile is sufficiently ``curved"
(i.e., its second spatial derivative is big enough) and associate the parameter $B_0$
with a typical value of the field.

Assuming further electric conductivity and chemical composition to be constant,
$\sigma_0= 5\times 10^{26}$s$^{-1}$ and the relative atomic weight $A/Z=25$, respectively,
we find the normalization field at a density $\rho = 10^{14}$g cm$^{-3}$
to be $7\times 10^{10}$G \citep[see, e.g.,][]{PGZ00}. That is, for typical (inner) crustal
magnetic 
fields ranging from $7\times 10^{12}$G to $1.4\times 10^{14}$G we find 
a $B_0$ between
$100$ and $2000$ and the e--folding time of the most rapidly growing unstable
mode to be $0.0035$ and $0.0003$ times the ohmic decay time, respectively.
Note, when comparing with the values of Fig.~\ref{fig2} that the dimensionless
thickness 2 of the slab causes a factor 4 in the growth times.  
These values agree roughly with the ratios $\tau_B/\tau_{\text{ohm}}$ derived in Sect.~\ref{obsmot}
from observational data.  
An initial perturbation will 
quickly evolve to a level at which the linear analysis is no longer feasible,
that is, at which it starts to drain a remarkable amount of energy out of the
background field.
Note, that the assumption about $B_0$ may well be in agreement with 
the surface field data given in Table~\ref{table}. 
 
We want to emphasize again that a sufficient curvature of the background field
profile
is a necessary condition for the occurrence of an unstable behavior. Therefore 
neither the derivation of the well--known helicoidal waves (whistlers) nor its
modification presented in \citet{VCO00} could reveal it because a homogeneous
background field was used there.

\begin{figure}[t]
\begin{center}
\hspace*{-5mm}\epsfig{file=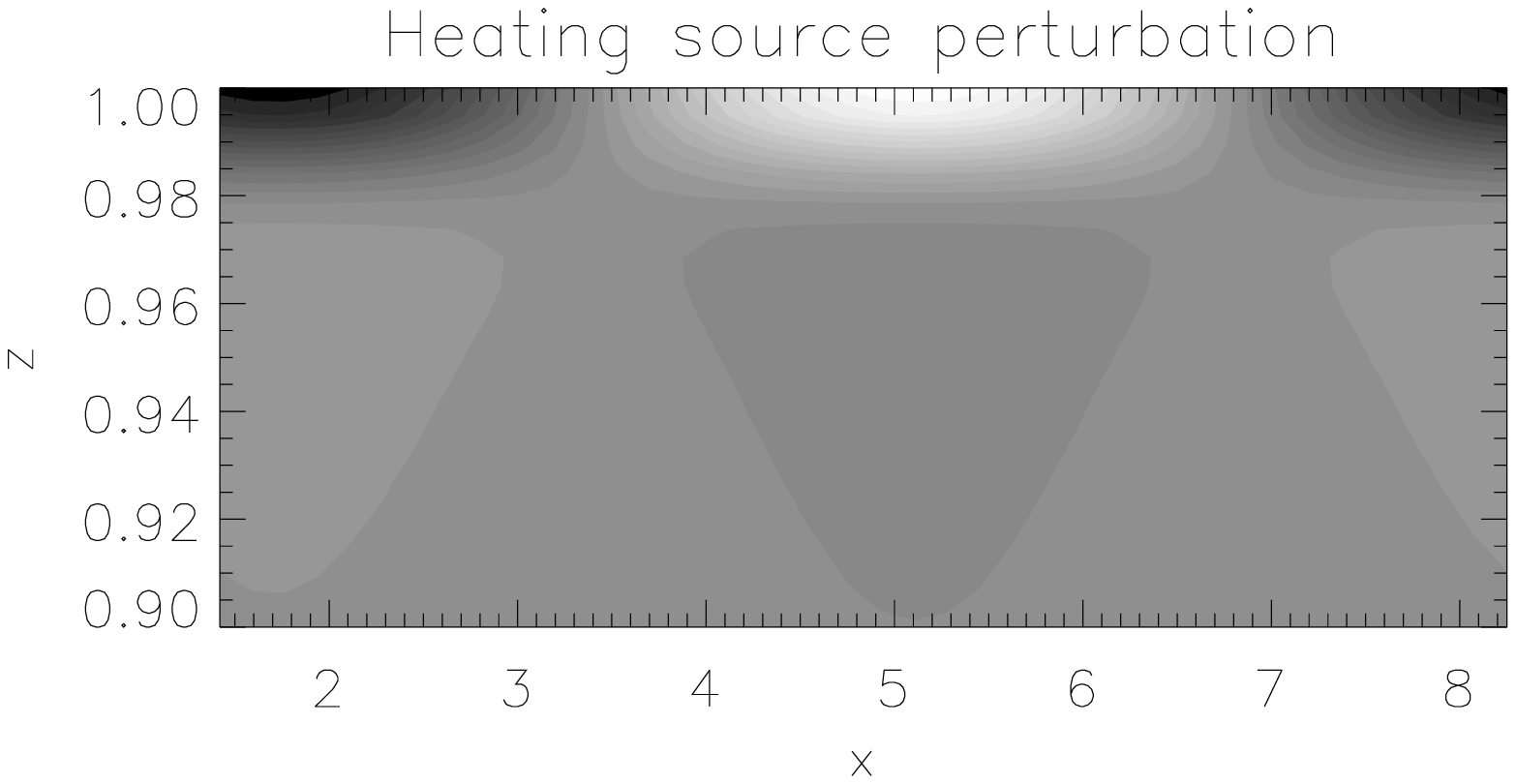, width=1.1\linewidth,clip=}
\end{center}

\vspace{-1.2cm}
\begin{center}
\hspace*{-5mm}\epsfig{file=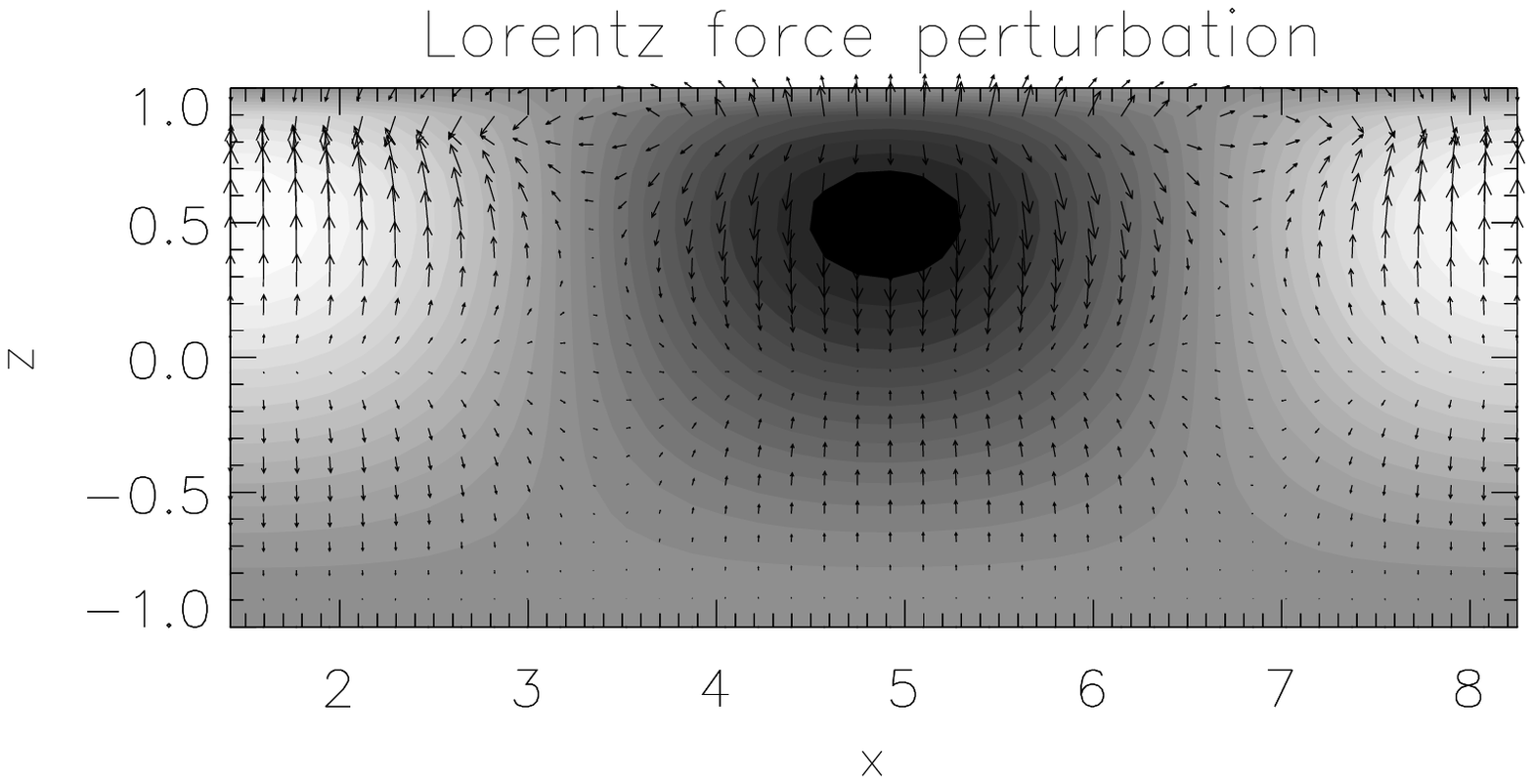, width=1.1\linewidth,clip=}
\end{center}

\vspace{-.3cm}
\caption{Heat source density perturbation $2\curl\Bvec_0\cdot\curl\bvec$ (upper panel) and
Lorentz force density perturbation $\curl\Bvec_0\times\bvec + \curl\bvec\times\Bvec_0$
(lower panel) of the fastest growing mode for $B_0=2000$ in arbitrary units.
Upper panel: dark and light shading denotes cooling and heating by the perturbation, respectively; note the stretched $z$--scale. 
Lower panel: arrows denote $x$-- and $z$--force components, grey shading the value of the $y$--component (dark -- into, light -- out
of the plane).}
\label{fig7}
\end{figure}

With great reserve we may speculate about possible observational consequences.
For SOINSs, the small--scale field modes initially generated or existing
in the crust have already been decayed and the
magnetic field is concentrated almost completely in the large--scale, say,
dipolar mode.
Simultaneously, in the process of cooling the coefficient $m_e^* c/ e\tau_e$, i.e.,
the normalization field $B_N$, becomes smaller and smaller until the Hall--term in
\Eqref{equ:indeqdimless} dominates the linear ohmic term. Of course, as a 
counteracting tendency, the magnetic field decay will continue, increasingly modified
by the Hall--cascade and the gradual onset of the Hall--instability. Therefore, we 
are confronted with a competition of one process increasing $B_0$ (via $\tau_e$)
and another diminishing it (via the {\em non--normalized} magnetic field). The
final decision whether $B_0$ indeed reaches values $O(1000)$ or even higher has to be left
to non--linear coupled magneto--thermal calculations.       
Once the Hall--term dominates the instability may
raise small--scale perturbations down to scale lengths $\gtrsim l_{\text{crit}}$ on expense of
the dipolar mode, resulting in a
spin--down behaviour of SOINSs as discussed in Sect.~\ref{obsmot}.

Another possible observational consequence
follows from the dissipative effects of the rapidly excited small--scale field modes,
that is, enhanced Joule heating.
As can be seen in the upper panel of Fig.~\ref{fig7}, the growing perturbations cause
heat sources located close
to the surface of the crust which, in turn, may produce a patchy neutron star surface.
Note, that the plot shows only the part linear in $\bvec$ of the heat source distribution,
since the quadratic term $(\curl\bvec)^2$ has to be dropped
in the linear analysis. 
These heat sources are concentrated in a very thin layer comprising only less than 5 \%
of the crustal thickness.
Clearly, conclusions about the observability of that hot spots are impossible to be drawn
from our simple model, since to infer the temperature distribution from the heat sources
demands the knowledge of the absolute perturbation amplitudes.
Anyway, we expect the neutron star's surface to be warmer during the episode of the Hall--instability
than standard cooling calculations predict. 

Third, strong small--scale field components cause small--scale Lorentz forces.
The lower panel of Fig.~\ref{fig7} shows that the maximum of the Lorentz force density 
perturbation is concentrated in
deeper regions of the crust (at about 25 \% of the thickness). From its pattern one may infer shear,
both in $y$-- and $z$-- directions, to be the prevailing type of deformation. 
\citet{TD95} discussed a scenario in
which the helicoidal waves, being a consequence of the Hall--drift, ignite high energetic
bursts by cracking the crust. From our results we can hope that the unstable modes of the Hall--instability
act even more efficiently than the at best undamped, but never growing helicoidal waves can.     

However, all the questions connected with the actual influences of the presented instability
on spin--down, cooling and crust--cracking can be decided only by
performing non--linear calculations in a realistic spherical--shell geometry
taking into account a realistic density profile and chemical composition 
of the crust.
On the other hand, the question at which field strength the action of the Hall--instability
finally ceases, that is, the field decay is again overtaken by ohmic dissipation and, as a consequence, 
to which extent the distribution of observable pulsars in a $B$--$P$, better, in a
$B$--$P$--$\tau_{\text{a}}$ diagram is possibly influenced, can again only be answered
on the basis of realistic non--linear magneto--thermal simulations.
Whether the scenario of an accelerated field decay is in agreement 
with the observed properties of the pulsar population can only be decided
by help of a population synthesis as presented by \citet{RF01}
applying a decay pattern with an initially slow decay followed by an episode
of rapid decay and leading finally to a nearly constant field.

\begin{acknowledgements}
We are grateful to R. Hollerbach, G. R\"udiger and A. Schwope for stimulating
discussions.
\end{acknowledgements}

\bibliographystyle{aa}
\bibliography{Hall3}

\begin{thebibliography}{34}
\expandafter\ifx\csname natexlab\endcsname\relax\def\natexlab#1{#1}\fi

\bibitem[{Becker {et~al.}(1996)Becker, Brazier, \& Tr\"umper}]{BBT96}
Becker, W., Brazier, K., \& Tr\"umper, J. 1996, \aap, 306, 464

\bibitem[{Br\"auer \& R\"adler(1988)}]{BR88}
Br\"auer, H.-J. \& R\"adler, K.-H. 1988, Astron. Nachr., 309, 1

\bibitem[{Casini \& Montemayor(1998)}]{CM98}
Casini, H. \& Montemayor, R. 1998, \apj, 503, 374

\bibitem[{Geppert {et~al.}(1999)Geppert, Page, \& Zannias}]{GPZ99}
Geppert, U., Page, D., \& Zannias, T. 1999, \aap, 345, 847

\bibitem[{Goldreich \& Reisenegger(1992)}]{GR92}
Goldreich, P. \& Reisenegger, A. 1992, \apj, 395, 250

\bibitem[{Haensel {et~al.}(1990)Haensel, Urpin, \& Yakovlev}]{HUY90}
Haensel, P., Urpin, V., \& Yakovlev, D. 1990, \aap, 229, 133

\bibitem[{Hollerbach \& R\"udiger(2002)}]{HR02}
Hollerbach, R. \& R\"udiger, G. 2002, \mnras, submitted

\bibitem[{Johnston \& Galloway(1999)}]{JG99}
Johnston, S. \& Galloway, D. 1999, \mnras, 306, L50

\bibitem[{Krolik(1991)}]{K91}
Krolik, J. 1991, \apjl, 373, L69

\bibitem[{Kundt(2001)}]{K01}
Kundt, W. 2001, Bull. Astr. Soc. India, 29, 283

\bibitem[{Lyne {et~al.}(1996)Lyne, Pritchard, Graham-Smith, \& Camilo}]{LPGC96}
Lyne, A., Pritchard, R., Graham-Smith, F., \& Camilo, F. 1996, Nature, 381, 497

\bibitem[{{\mbox{Manchester et al.}}(2001)}]{M01}
{\mbox{Manchester et al.}} 2001,
  http://www.atnf.csiro.au/people/pulsar/catalogue/

\bibitem[{Melatos(1997)}]{M97}
Melatos, A. 1997, \mnras, 288, 1049

\bibitem[{Muslimov(1994)}]{M94}
Muslimov, A. 1994, \mnras, 267, 523

\bibitem[{Naito \& Kojima(1994)}]{NK94}
Naito, T. \& Kojima, Y.~. 1994, \mnras, 266, 597

\bibitem[{Page {et~al.}(2000)Page, Geppert, \& Zannias}]{PGZ00}
Page, D., Geppert, U., \& Zannias, T. 2000, \aap, 360, 1052

\bibitem[{R\"adler(1969)}]{R69}
R\"adler, K.-H. 1969, Monatsber. Dt. Akad. Wiss., 11, 194

\bibitem[{Regimbau \& de~Freitas~Pacheco(2001)}]{RF01}
Regimbau, T. \& de~Freitas~Pacheco, J. 2001, \aap, 374, 182

\bibitem[{Rheinhardt \& Geppert(2002)}]{RG02}
Rheinhardt, M. \& Geppert, U. 2002, \prl, 88, 101103

\bibitem[{Shalybkov \& Urpin(1995)}]{SU95}
Shalybkov, D. \& Urpin, V. 1995, \mnras, 273, 643

\bibitem[{Shalybkov \& Urpin(1997)}]{SU97}
---. 1997, \aap, 321, 685

\bibitem[{Shapiro \& Teukolsky(1983)}]{ST83}
Shapiro, S. \& Teukolsky, S. 1983, Black Holes, White Dwarfs and Neutron Stars
  (New York: John Wiley \& Sons)

\bibitem[{Shibazaki \& Hirano(1995)}]{SH95}
Shibazaki, N. \& Hirano, S. 1995, in Annals of New York Academy of Sciences,
  Vol. 759, Seventeenth Texas Symposium on Relativistic Astrophysics and
  Cosmology, 295

\bibitem[{Shibazaki \& Mochizuki(1995)}]{SM95}
Shibazaki, N. \& Mochizuki, Y. 1995, \apj, 438, 288

\bibitem[{Tauris \& Konar(2001)}]{TK01}
Tauris, T. \& Konar, S. 2001, \aap, 376, 543

\bibitem[{Taylor {et~al.}(1993)Taylor, Manchester, \& Lyne}]{TML93}
Taylor, J., Manchester, R., \& Lyne, A. 1993, \apjs, 88, 529

\bibitem[{Thompson \& Duncan(1995)}]{TD95}
Thompson, C. \& Duncan, R. 1995, \apj, 473, 322

\bibitem[{Urpin {et~al.}(1986)Urpin, Levshakov, \& Yakovlev}]{ULY86}
Urpin, V., Levshakov, S., \& Yakovlev, D. 1986, \mnras, 219, 703

\bibitem[{Urpin \& Shalybkov(1995)}]{US95}
Urpin, V. \& Shalybkov, D. 1995, \mnras, 294, 117

\bibitem[{Urpin \& Shalybkov(1999)}]{US99}
---. 1999, \mnras, 304, 451

\bibitem[{Urpin \& Yakovlev(1980)}]{UY80}
Urpin, V. \& Yakovlev, D. 1980, Sov. Astron., 24, 425

\bibitem[{Vainshtein {et~al.}(2000)Vainshtein, Chitre, \& Olinto}]{VCO00}
Vainshtein, S., Chitre, S., \& Olinto, A. 2000, \pre, 61, 4422

\bibitem[{Wiebicke \& Geppert(1996)}]{WG96}
Wiebicke, H.-J. \& Geppert, U. 1996, \aap, 309, 203

\bibitem[{Yakovlev \& Shalybkov(1991)}]{YS91}
Yakovlev, D. \& Shalybkov, D. 1991, \apss, 176, 191

\end{thebibliography}
\end{document}